\documentclass[conference,10pt]{IEEEtran}
\usepackage{amsmath,graphicx}
\usepackage{pgf,tikz,pgfplots}

\usepackage{amsmath,amssymb,amsthm}
\usepackage{algorithmic}
\usepackage[linesnumbered,boxed]{algorithm2e}
\usepackage{bm}
\usepackage{color}

\usepackage{caption}

\usepackage{cases}

\usepackage{subfigure}

\newcommand{\rev}[1]{{\color{black}{#1}}}

\newcommand{\blue}[1]{{\color{blue}{#1}}}

\def\simpleFlowConservation{1}

\def\notvartualgraph{1}
\ifthenelse{\notvartualgraph=1}{
	\def\barr#1{{{#1}}}
	\def\xhatzero{x}
}
{
	\def\barr#1{{\bar{#1}}}
	\def\xhatzero{x^0}
}

\begin{document}
	
\title{Network Slicing for Service-Oriented Networks with Flexible Routing and Guaranteed E2E Latency\thanks{\!\!\!\rev{Corresponding author: Ya-Feng Liu.}}}
\author{\IEEEauthorblockN{Wei-Kun Chen\IEEEauthorrefmark{1},
		Ya-Feng Liu\IEEEauthorrefmark{2},
		Antonio De Domenico\IEEEauthorrefmark{3},
		{Zhi-Quan Luo\IEEEauthorrefmark{4}}
	}
	\IEEEauthorblockA{\IEEEauthorrefmark{1}School of Mathematics and Statistics/Beijing Key Laboratory on MCAACI, Beijing Institute of Technology, Beijing, China}
	\IEEEauthorblockA{\IEEEauthorrefmark{2}LSEC, ICMSEC, AMSS, Chinese Academy of Sciences, Beijing, China}
	\IEEEauthorblockA{\IEEEauthorrefmark{3}CEA-LETI, Grenoble, France}
		\IEEEauthorblockA{\IEEEauthorrefmark{4}Shenzhen Research Institute of Big Data and The Chinese University of Hong Kong, Shenzhen, China}
	\small{Email: chenweikun@bit.edu.cn, yafliu@lsec.cc.ac.cn {(Corresponding author),} antonio.de-domenico@cea.fr, luozq@cuhk.edu.cn}}

\maketitle

\begin{abstract}
	Network function virtualization is a promising technology to simultaneously support multiple services with diverse characteristics and requirements in the fifth generation and beyond networks.
	In practice, each service consists of a predetermined sequence of functions, called service function chain (SFC), running on a cloud environment.
	To make different service slices work properly in harmony, it is crucial to select the cloud nodes to deploy the functions in the SFC and flexibly route the flow of the services such that these functions are processed in sequence, the end-to-end (E2E) latency constraints of all services are guaranteed, and all resource constraints are respected.
	In this paper, we propose a new (mixed binary linear program) formulation of the above network slicing problem that {optimizes the system energy efficiency while} jointly considers the resource budget, functional instantiation, flow routing, and E2E latency requirement.
	Numerical results show the advantage of the proposed formulation compared to the existing ones.
\end{abstract}


\section{Introduction}

Network function virtualization (NFV) is considered as one of the key technologies for the fifth generation (5G) and beyond 5G (B5G) networks \cite{Mijumbi2016}.
In contrast to traditional networks where service functions are processed by dedicated hardwares in fixed locations, NFV can efficiently take the advantage of cloud technologies to configure some specific nodes in the network to process network service functions on-demand, and then flexibly establish a customized virtual network for each service request.
In the NFV-enabled network, classic networking nodes are integrated with NFV-enabled nodes (i.e., cloud nodes) and each service consists of a predetermined sequence of virtual network functions (VNFs), called service function chain (SFC) \cite{Zhang2013,Mirjalily2018}, which can only be processed by certain specific cloud nodes {\cite{Zhang2017}}.
In practice, each service flow has to pass all VNFs in sequence and its end-to-end (E2E) latency requirement must be satisfied.
However, since all VNFs run over a shared common network infrastructure, it is crucial to allocate cloud and communication resources to meet the diverse service requirements, subject to the SFC constraints and the E2E latency constraints of all services and all cloud nodes' and links' capacity constraints.

The above resource allocation problem in the NFV-enabled network is called \emph{network slicing} in the literature and considerable works have been done on it recently; see \cite{Zhang2017}-\cite{Charikar2019} 
and the references therein.
More specifically, references \cite{Zhang2017,Zhang2019} considered the VNF deployment problem with a limited network resource constraint, but it did not take the E2E latency constraint of each service into consideration.
Reference \cite{Domenico2019} investigated a specific two-layer network which consists of a  {central} cloud node and several edge cloud nodes without considering the limited link capacity, which may lead to violations of resource constraints.
{Reference \cite{Jiang2012} simplified the routing strategy by selecting paths from a predetermined path set.}
%
{Reference \cite{Addis2015} simplified the VNF placement decision-making by assuming that all VNFs in a SFC must be instantiated at the same cloud node.}
Reference \cite{Woldeyohannes2018} proposed a way of analyzing the dependencies between traffic routing and VNF placement in the NFV networks. Though the E2E latency requirement of each service was integrated into the formulation, only a single path was allowed to route the traffic flow of each service.
Apparently, such a formulation does not fully exploit the flexibility of traffic routing and hence might affect the performance of the whole network.
Reference \cite{Charikar2019} assumed that instantiation of a VNF can be split over multiple cloud nodes, which may result in high coordination overhead in practice.

To the best of our knowledge, for the network slicing problem, none of the existing formulations/works simultaneously takes all of the above practical factors (e.g., flexible routing, E2E latency, and coordination overhead) into consideration.
The goal of this work is to provide a mathematical formulation of the network slicing problem that simultaneously allows the traffic flows to be flexibly transmitted on (possibly) multiple paths, satisfies the E2E latency requirements of all services, and requires that each service function in a SFC is processed by exactly one cloud node.
In particular, we formulate the problem as a novel mixed binary \emph{linear} program (MBLP), which minimizes the total power consumption of the whole network subject to the SFC constraints and the E2E latency constraints of all services and all cloud nodes' and links' capacity constraints.
Numerical results show the effectiveness of our proposed formulation.
\section{System model and problem formulation}
\label{sec:modelformulation}
In this section, we give a MBLP formulation for the network slicing problem.
\subsection{System Model}
Consider a communication network $\mathcal{G}=\{\mathcal{I},\mathcal{L}\}$, where $\mathcal{I}=\{i\}$ is the set of nodes and $\mathcal{L}=\{(i,j)\}$ is the set of links.
The network supports a set of flows $\mathcal{K}=\{k\}$.
We assume that each link $ (i,j) $ has an expected delay $ d_{ij} $ \cite{Woldeyohannes2018}, and a total data rate upper bounded by the capacity $C_{ij}$.
Let $ \mathcal{V} $ be a subset of $ \mathcal{I} $ denoting the set of the cloud nodes.
Each cloud node $ v $ has a computational capacity $ \mu_v $ and we assume as in \cite{Zhang2017} that processing one unit of data rate requires one unit of (normalized) computational capacity.
Let $S(k)$ and $D(k)$ be the source and destination of flow $k$, respectively, and suppose that $S(k),D(k)\notin \mathcal{V}$.
Each flow $ k $ relates to a distinct service, which is given by a SFC consisting of $ \ell_k $ functions that have to be performed in sequence by the network:
\begin{equation}
	\label{sequence}
	f_{1}^k\rightarrow f_{2}^k\rightarrow \cdots \rightarrow f_{\ell_k}^k.
\end{equation}
As required in \cite{Zhang2017,Domenico2019}, and \cite{Woldeyohannes2018}, to minimize the coordination overhead, each function must be instantiated at exactly one cloud node.
If function $ f^k_s $, $ s \in \mathcal{F}(k) := \{1,\ldots, \ell_k\} $, is processed by node $ v $ in $ \mathcal{V} $, we assume the expected NFV delay is known as $ d_{v,s}(k) $ which includes both processing delay and queuing delay, as in \cite{Woldeyohannes2018}.
For flow $ k $, denote $ \lambda_0(k) $ and $ \lambda_s(k) $ as the service function rates before receiving any {function} and after receiving {function} $ f^k_s $, respectively. Each flow $ k $ is required to have an {E2E latency} guarantee, denoted as $ \Theta_k $.
\subsection{Preview of the New Formulation}
The network slicing problem is to determine functional instantiation of all flows and the routes and {associated} data rates of all flows on the routes while satisfying the SFC requirements, the E2E delay requirements, and the capacity constraints on all cloud nodes and links.
In this section, we shall provide a new problem formulation of the network slicing problem which takes practical factors like coordination overhead, flexible routing, and E2E latency requirements into consideration; see Eq. \eqref{mip} further ahead.

Our proposed formulation builds upon those in two closely related works \cite{Woldeyohannes2018} and \cite{Zhang2017} but takes further steps.
More specifically, in sharp contrast to the formulation in \cite{Woldeyohannes2018} where only a \emph{single} path is allowed to route the traffic flow of each service (between two cloud nodes processing two adjacent functions of a service), our proposed formulation allows the traffic flow of each service to transmit on (possibly) multiple paths and hence fully exploits the flexibility of traffic routing; different from that in \cite{Zhang2017}, our formulation guarantees the E2E delay of all services, which consists of two types of delays: total communication delay on the links and total NFV delay on the cloud nodes.
Next, we describe the constraints and objective function of our formulation in details.
\subsection{VNF Placement and Node Capacity Constraints}
\label{subsec:constraint}
We introduce the binary variable $x_{v,s}(k),~s=1,\ldots,\ell_k$, to indicate whether or not node $v$ {in $\mathcal{V}$ processes function $f^k_s$}, i.e.,
\begin{eqnarray*}
	x_{v,s}(k)&=&\left\{\begin{array}{ll}1,
		   & {\text{if~node}}~ v ~{\text{{processes}~function}}~f^k_s;~ \\
		0, & {\text{otherwise}}.\end{array}\right.
\end{eqnarray*}
%
%
For simplicity of presentation as in \cite{Zhang2017}, we require that each cloud node processes at most one function for each flow:
\begin{equation}\label{key}
	\sum_{s \in \mathcal{F}(k)} x_{v,s} (k) \leq 1, \ \forall~k\in \mathcal{K}, \ \forall~v \in \mathcal{\barr{V}}.
\end{equation}

For each flow $ k $, we require that each service function in the chain $\mathcal{F}(k)$ is served by exactly one cloud node, i.e.,
\begin{eqnarray}
	\label{onlyonenode}
	\sum_{v\in \mathcal{V}}\xhatzero_{v,s}(k)=1,~\forall ~k \in \mathcal{K},~ \forall ~s\in  \mathcal{F}(k).\label{3}
\end{eqnarray}
Since processing one unit of data rate consumes one unit of (normalized) computational capacity, we can get the node capacity constraints as follows:
\begin{eqnarray}
	\sum_{k\in \mathcal{K}}\sum_{s \in \mathcal{F}(k)}\lambda_s(k)\xhatzero_{v,s}(k)\leq \mu_v,~\forall~ v \in \mathcal{V}.\label{13}
\end{eqnarray}
Let $y_v\in\{0,1\}$ represent the activation of cloud node $v$, i.e., if $ y_v =1 $, node $ v $ is activated and powered on; otherwise, it is powered off. Thus
\begin{eqnarray}
	\xhatzero_{v,s}(k) \leq  y_v, ~ \forall~v \in \mathcal{V}, ~\forall~k \in \mathcal{K}, ~\forall~s \in \mathcal{F}(k). \label{2}
\end{eqnarray}
\subsection{Flexible Routing and Link Capacity Constraints}
We assume that there are at most $ P $ paths between any pair of cloud nodes that processes two adjacent functions of a flow.
In general, such an assumption on the number of paths may affect the solution's quality.
Indeed, the choice of  $P$ offers a tradeoff between the flexibility of routing in the problem formulation and the computational complexity of solving it: the larger the parameter $P$ is, the more flexibility of routing and the higher the computational complexity.
%
%

Denote $ \mathcal{P}=\{1, \ldots,P\} $.
%
If {cloud nodes} $ v_s $ and $ v_{s+1} $ are used to host {the $ s $-th and $ (s+1) $-th functions of flow $ k $ (i.e., functions $ f^k_s $ and $ f^k_{s+1} $)}, respectively, and path $ p $ is used to route the traffic flow, let $ r(k,s,v_s, v_{s+1}, p) $ be the associated amount of the data rate.
We need to introduce this variable in our formulation, as the traffic flow of each service in our formulation is allowed to transmit on (possibly) multiple paths in order to exploit the flexibility of traffic routing. 
%
%
In particular, if $ s=0 $, we assume $ v_s= S(k) $ and if $ s=\ell_k $, we assume $ v_{s+1} = D(k) $.
Notice that by \eqref{key} and the fact that $ S(k), D(k) \notin \mathcal{V} $, we must have $ v_s \neq v_{s+1} $.
For each $ k \in \mathcal{K} $, from the definitions of $ x_{v_s,s}(k) $, $ x_{v_{s+1}, s+1}(k) $, and $ r(k,s,v_s, v_{s+1}, p) $, we have
	\begin{align}
		 & \sum_{p \in \mathcal{P}}  r(k, {s}, v_s, v_{s+1}, p) =  \lambda_{s}(k) x_{v_s,s}(k)  x_{v_{s+1},{s+1}}(k),   \nonumber                      \\[-5pt]
		 & {\qquad \qquad\qquad\qquad \forall~s\in \mathcal{F}(k)\backslash\{\ell_k\},~\forall~v_s, v_{s+1} \in \mathcal{\barr{V}}. \label{relalambdaandx1}}
	\end{align}
Constraint \eqref{relalambdaandx1} indicates that if the $ s $-th and $ (s+1) $-th functions of flow $ k $ (i.e., functions $ f_s^k $ and $ f_{s+1}^k $) are hosted at cloud nodes $ v_s $ and $ v_{s+1} $, respectively, then the sum of all the {data} rates sent from $ v_s $ to $ v_{s+1} $ must be equal to $ \lambda_s(k) $.
\ifthenelse{\simpleFlowConservation=0}{
	\blue{Similarly, if function $ f_1^k $ is hosted at cloud node $ v_1 $, constraint \eqref{relalambdaandx2} guarantees that the sum of all the {data} rates sent from $S(k) $ to $ v_{1} $ must be equal to $ \lambda_0(k) $; if function $ f_{\ell_k}^k $ is hosted at {cloud} node $ v_{\ell_k} $, constraint \eqref{relalambdaandx3} guarantees that the sum of all the data rates sent from $v_{\ell_k} $ to $ D(k) $ must be equal to $ \lambda_{\ell_k}(k) $.}
}
We then use $ z_{ij}(k, s,v_s, v_{s+1}, p) =1 $ to denote that the $ s $-th and $ (s+1) $-th functions of flow $ k $ (i.e., functions $ f_{s}^k $ and $ f_{s+1}^{k} $) are processed by cloud nodes $ v_s $ and $ v_{s+1} $, respectively, path $ p $ is used to route the associated traffic flow, and link $ (i,j) $ is on path $ p $; otherwise, $ z_{ij}(k, s,v_s, v_{s+1}, p) =0 $.
By definition, for all $ k \in \mathcal{K} $, $ p \in \mathcal{P} $, and $ (i,j) \in \mathcal{\barr{L}} $, we have
%
\begin{align}
		 & z_{ij}(k, s, v_s,v_{s+1}, p ) \leq  x_{v_s,s}(k) x_{v_{s+1}, {s+1}}(k),\nonumber                                                            \\
		 & \qquad \qquad\qquad\qquad\forall~s \in \mathcal{F}(k)\backslash\{\ell_k\}, ~\forall~v_s, v_{s+1} \in \mathcal{\barr{V}}. \label{relazandx1}
\end{align}
%
If $ z_{ij}(k, s,v_s, v_{s+1}, p) =1 $, let $ r_{ij}(k, s,v_s, v_{s+1}, p )  $ denote the associated amount of {data} rate.
By definition, for each $ k \in \mathcal{K} $, $ p\in \mathcal{P} $, and $ (i,j) \in \mathcal{\barr{L}} $, we have the following coupling constraints:
	\begin{align}
		 & r_{ij}(k, s, v_s, v_{s+1}, p ) \leq \lambda_{s}(k)  z_{ij}(k, s, v_s, v_{s+1},p )\nonumber,                                                 \\
		 & \qquad \qquad\qquad\qquad\forall~s \in \mathcal{F}(k)\backslash\{\ell_k\}, ~\forall~v_s, v_{s+1} \mathcal{\in \barr{V}}. \label{relarandz1}
	\end{align}
%
The total data rates on link $ (i,j) $ is upper bounded by capacity $ C_{ij} $:
\begin{eqnarray}
	\sum_{k \in \mathcal{K}} \sum_{p \in \mathcal{P}} \sum_{s\in \mathcal{F}(k) \cup \{0\}} \sum_{v_s\in \mathcal{\barr{V}}}  \sum_{v_{s+1}\in \mathcal{\barr{V}}\backslash\{v_s\}} r_{ij}(k, s, v_s, v_{s+1}, p)\nonumber\\
	\leq C_{ij}, ~  \forall~(i,j) \in \mathcal{L}\label{1} .
\end{eqnarray}
\subsection{SFC Constraints}
To ensure the functions of each flow {are} followed in the prespecified order as in \eqref{sequence}, we need to introduce several constraints below.
%
For each $ k \in \mathcal{K} $, $ p \in \mathcal{P} $, $ v_s, v_{s+1}  \in \mathcal{\barr{V}}$ with $ s \in \mathcal{F}(k)\backslash\{\ell_k\} $, and $ i \in \mathcal{\barr{I}} $, we have
\begin{equation*}
\sum_{j: (j,i) \in \mathcal{\barr{L}}} r_{ji}(k, s, v_s, v_{s+1}, p) - \sum_{j: (i,j) \in \mathcal{\barr{L}}} r_{ij}(k, s, v_s, v_{s+1}, p)=
\end{equation*}
\begin{numcases}{}
-r(k, s, v_s, v_{s+1},p),     & ~~~\text{if}~$i = v_s;$ \label{mediacons1} \\
0,                                   & ~~~\text{if}~$i \neq v_s, ~v_{s+1};$  \label{mediacons2}   \\
r(k, s, v_s, v_{s+1},p),    &  ~~~\text{if}~$i=v_{s+1};$  \label{mediacons3}   
\end{numcases}
\begin{equation*}
\sum_{j: (j,i) \in \mathcal{\barr{L}}} z_{ji}(k, s, v_s, v_{s+1}, p) - \sum_{j: (i,j) \in \mathcal{\barr{L}}} z_{ij}(k, s, v_s, v_{s+1}, p)=
\end{equation*}
\begin{numcases}{}
-x_{v_s,s}(k)  x_{v_{s+1},{s+1}}(k),     & \text{if}~$i = v_s;$\label{mediacons4}   \\
0,                                   & \text{if}~$i \neq v_s, ~v_{s+1};$ \label{mediacons5}  \\
x_{v_s,s}(k)  x_{v_{s+1},{s+1}}(k)    &  \text{if}~$i=v_{s+1}$.    \label{mediacons6}
\end{numcases}  
{First}, note that constraints \eqref{mediacons1}, \eqref{mediacons2}, and \eqref{mediacons3} are flow conservation constraints for the data rate.
Second, we need another three flow conservation constraints \eqref{mediacons4}, \eqref{mediacons5}, and \eqref{mediacons6}.
To be more precise, for each pair of cloud nodes $ v_s $ and $ v_{s+1} $, considering constraints \eqref{mediacons4}, \eqref{mediacons5}, and \eqref{mediacons6}, we only need to look at the case that $ x_{v_s,s}(k) =1 $ and $ x_{v_{s+1}, s+1}(k)=1 $ since otherwise from constraint \eqref{relazandx1}, all the variables $ z_{ij}(k,s,v_s,v_{s+1}, p) $ in \eqref{mediacons4}, \eqref{mediacons5}, and \eqref{mediacons6} must be equal to zero.
Constraint \eqref{mediacons5} enforces that for every node that does not host functions $ f_s^k $ and $ f_{s+1}^k $, if one of its incoming links is used to route the flow from node $ v_s $ to node $ v_{s+1} $ on path $ p $, then one of its outgoing links is also assigned for this path.
Similarly, constraints \eqref{mediacons4} and \eqref{mediacons6} imply that, if function $ f_s^k $ is hosted at node $ v_s $, one of the outgoing links of node $ v_s $ must be assigned for path $ p $ and, if function $ f_{s+1}^k $ is hosted at node $ v_{s+1} $, one of the incoming links of node $ v_{s+1} $ must be assigned for path $ p $, respectively.

%
%
%
\subsection{E2E Latency Constraints}
Next, we consider the delay constraints.
Let $ \theta(k,s,{s+1}) $ be the variable denoting the communication delay due to the traffic flow from the cloud node hosting function $ f^k_s $ to the cloud node hosting function $ f^k_{s+1} $.
Then, $ \theta(k,s,{s+1}) $ should be the largest one among the $ P $ paths, i.e.,
\begin{align}
	 & \theta(k,s,{s+1}) \geq \sum_{v_s,v_{s+1}\in\mathcal{\barr{V}}} \sum_{(i,j) \in \mathcal{L}}  d_{ij}  z_{ij}(k, s, v_s, v_{s+1}, p), \nonumber \\
	 & \qquad \qquad\forall ~p \in \mathcal{P}, ~ \forall~k \in \mathcal{K}, ~ \forall~s \in \mathcal{F}(k) \cup \{0\} \label{consdelay2funs}.
\end{align}
Hence the total communication delay on the links of flow $ k $, denoted as $ \Theta_L(k) $, can be written as
\begin{equation}
	\Theta_L(k) = \sum_{s \in \mathcal{F}(k)\cup \{0\}} \theta(k,s,{s+1}), ~\forall~ k \in  \mathcal{K}.
\end{equation}
Now for each flow $ k $, we consider the total NFV delay on the nodes, denoted as $ \Theta_N(k) $.
This can be written as
\begin{equation}
	\Theta_N(k) = \sum_{s \in \mathcal{F}(k)} \sum_{v \in \mathcal{{V}}} d_{v,s}(k) \xhatzero_{v,s}(k),~\forall  ~k \in  \mathcal{K}.
\end{equation}
{The E2E delay of flow $k$ is {the} sum of total communication delay $ \Theta_L(k) $ and total NFV delay $ \Theta_N(k) $.}
The following delay constraint ensures that {flow $k$'s} {E2E} delay is less than or equal to its threshold $\Theta_k$:
\begin{equation}
	\label{delayconstraint}
	\Theta_L(k)+\Theta_N(k) \leq \Theta_k,~\forall~k \in  \mathcal{K}.
\end{equation}
\subsection{A New MBLP Formulation}
\label{subsec:mipformulation}
The power consumption of a cloud node is the combination
of the dynamic load-dependent power consumption
(that increases linearly with the load) and the static power
consumption \cite{3gpp}.
Our objective is to minimize the total power consumption of the whole network:
\begin{align*}
	\sum_{v \in \mathcal{V}}\left[\beta_1y_v+{\Delta} \sum_{k \in \mathcal{K}}\sum_{s \in \mathcal{F}(k)} \lambda_s(k)x_{v,s}(k)\right] +\sum_{v \in \mathcal{V}}\beta_2(1-y_v)\label{objfuns}.
\end{align*}
In the above, the parameters $\beta_1$ and $\beta_2$ are the power consumptions of each activated cloud node and inactivated cloud node, respectively, satisfying $\beta_1>\beta_2$; the parameter $ \Delta $ is the power consumption of processing one unit of data rate.
From \eqref{onlyonenode}, the above objective function can be simplified as
	$(\beta_1-\beta_2)\sum_{v \in \mathcal{V}}y_v +c,$
where $  c=  \beta_2|\mathcal{V}|+{\Delta}\sum_{k \in \mathcal{K}}\sum_{s \in \mathcal{F}(k)} \lambda_s(k) $ is a constant.
Hence, minimizing the total power consumption is equivalent to minimizing the total number of activated cloud nodes. Based on the above analysis, we obtain the following problem formulation:
\begin{equation}
	\label{mip}
	\begin{aligned}
		 & \min_{\boldsymbol{x},\boldsymbol{y},\boldsymbol{z},\boldsymbol{r},\boldsymbol{\theta}} &  & \sum_{v \in \mathcal{V}}y_v, ~~~~~{\text{s.t.~}}                                                                                         &  & (\ref{key})-(\ref{delayconstraint}).
	\end{aligned}
\end{equation}
\subsection{Analysis Results of Problem \eqref{mip}}
We now present some analysis results of problem \eqref{mip} (without proofs due to the space reason).
First, problem \eqref{mip} is a MBLP (with both numbers of binary variables and linear constraints being ${\cal O}({|\mathcal{V}|}^2|\mathcal{L}||\mathcal{P}|\sum_{k \in \mathcal{K}}\ell_k )$) since the bilinear terms of binary variables $  x_{v_s,s}(k)  x_{v_{s+1},{s+1}}(k)  $ in \eqref{relalambdaandx1}, \eqref{relazandx1}, \eqref{mediacons5}, and \eqref{mediacons6} can be equivalently linearized. 
More specifically, we can replace the bilinear term $  x_{v_s,s}(k)  x_{v_{s+1},{s+1}}(k)$ by introducing an auxiliary binary variable $ \omega_{v_s, v_{s+1}}(k) $ and add the following linear constraints: $ \omega_{v_s, v_{s+1}}(k)  \leq x_{v_s,s}(k) $, $ \omega_{v_s, v_{s+1}}(k)  \leq x_{v_{s+1},{s+1}}(k) $, and $ \omega_{v_s, v_{s+1}}(k)\geq  x_{v_s,s}(k) + x_{v_{s+1},{s+1}}(k)- 1 $.
%
Note that the linearity of all variables in problem \eqref{mip} is vital, which allows to leverage the efficient integer programming solver such as Gurobi \cite{Gurobi} to solve the problem to global optimality.
%
%
%
Second, problem \eqref{mip} is strongly NP-hard.
Therefore, the above approach can only solve problem \eqref{mip} associated with small size networks.
In future works, we {shall develop} polynomial-time heuristic algorithms for solving problem \eqref{mip} to achieve the tradeoff between the performance and time complexity.
Third, 
if we set $P=1$ in \eqref{mip}, then our proposed formulation reduces to that in \cite{Woldeyohannes2018}.
In particular, the variables $ \boldsymbol{r}_{ij} $ in \eqref{1} can be replaced by those in the right-hand side of \eqref{relarandz1} and all constraints related to the variables $\boldsymbol{r}$ (e.g., \eqref{relalambdaandx1}, \eqref{relarandz1}, \eqref{mediacons1}, \eqref{mediacons2}, \eqref{mediacons3}) can be removed.
Our proposed formulation with $P>1$ allows the traffic flows to transmit over possibly multiple paths and fully exploits the flexibility of traffic routing.
\section{Numerical Results}
In this section, we present numerical results to illustrate the effectiveness of the proposed formulation.
\subsection{An Illustrative Example}
In this subsection, we show the performance of the proposed formulation by using an illustrative example.
\setlength{\intextsep}{5pt plus 2pt minus 2pt}
\begin{figure}[h]
	\captionsetup{belowskip=-4pt}
	\centering
	\begin{tikzpicture}[scale=\textwidth/23cm]
		\draw [->,line width=0.8pt] (1.6,5.4) -- (3.85,7.7);
		\draw [->,line width=0.8pt] (1.6,5.4) -- (3.95,3.05);
		\draw [->,line width=0.8pt] (3.93,7.77) -- (8.67,5.5);
		\draw [->,line width=0.8pt] (4,3) -- (8.7,5.43);
		\draw [->,line width=0.8pt] (4,3) -- (3.92,7.7);
		\draw [->,line width=0.8pt] (8.71,5.43) -- (6.05,5.43);
		\draw [->,line width=0.8pt] (5.97,5.35) -- (4,7.7);
		\begin{scriptsize}
			\draw [fill=black] (1.6,5.4) circle (2.5pt);
			\draw[color=black] (1.5,5.7) node {A};
			\draw [fill=black] (3.93,7.77) circle (2.5pt);
			\draw[color=black] (4.2,7.9) node {B};
			\draw [fill=black] (4,3) ++(-2.5pt,0 pt) -- ++(2.5pt,2.5pt)--++(2.5pt,-2.5pt)--++(-2.5pt,-2.5pt)--++(-2.5pt,2.5pt);
			\draw[color=black] (4.15,3.52) node {C};
			\draw[color=black] (4.5,3.48) node {(4)};
			\draw [fill=black] (5.97,5.43) circle (2.5pt);
			\draw[color=black] (6,5.7) node {D};
			\draw [fill=black] (8.71,5.43) ++(-2.5pt,0 pt) -- ++(2.5pt,2.5pt)--++(2.5pt,-2.5pt)--++(-2.5pt,-2.5pt)--++(-2.5pt,2.5pt);
			\draw[color=black] (8.8,5.7) node {E};
			\draw[color=black] (2.5,6.8) node {(2,1)};
			\draw[color=black] (3.2,4.3) node {(2,1)};
			\draw[color=black] (6.6,6.8) node {(2,1)};
			\draw[color=black] (6.1,4.4) node {(2,1)};
			\draw[color=black] (4.4,5.5) node {(2,1)};
			\draw[color=black] (7.3,5.6) node {(4,1)};
			\draw[color=black] (5.4,6.6) node {(2,1)};
			\draw[color=black] (9.15,5.67) node {(4)};
			\draw (6.5,3.48) node[anchor=north west] {{Cloud nodes}};
			\draw [fill=black] (6.3,3.3) ++(-2.5pt,0 pt) -- ++(2.5pt,2.5pt)--++(2.5pt,-2.5pt)--++(-2.5pt,-2.5pt)--++(-2.5pt,2.5pt);
		\end{scriptsize}
	\end{tikzpicture}
	\caption{{A toy network example where the pair $ (a,b) $ over each link denotes the link capacity $ a $ and the communication delay $ b $ and the value $ c $ inside the parentheses at each {cloud} node denotes the node capacity $ c $.}}
	\label{originalmap1}
\end{figure}
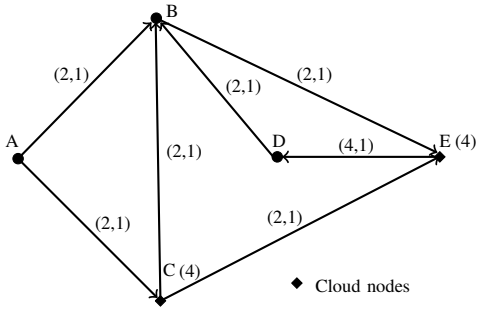

Consider the example in Fig. \ref{originalmap1}. There are two different functions available, i.e., $ f^1 $ and $ f^2 $. Cloud node C can only process function $ f^2 $, while cloud node E can {process} both functions $ f^1 $ and $ f^2 $.
Suppose there are two services where service I is from node A to node D with the E2E delay threshold $\Theta_{1}=4$ and service II is from node A to node B with the E2E delay threshold $\Theta_2=3$.
Functions $ f^1 $ and $ f^2 $ need to be processed for services I and II, respectively; for each service $ k $, the service function rates $ \lambda_0(k) $ and $ \lambda_1(k) $ are $1$; the NFV delays of both functions at (possible) cloud node C and cloud node E are $1$.
Solving problem \eqref{mip} {with $P=2$} gives the following solution:
$\begin{array}{l}
		\text{Service I}: \text{A} \rightarrow \text{B} \rightarrow \text{E~{(providing function $ f^1 $)}} \rightarrow  \text{D}, \\
		\text{Service II}: \text{A} \rightarrow  \text{C~{(providing function $ f^2 $)}} \rightarrow  \text{B}.
	\end{array}$\\
Note that since the communication delay on each link {$ (i,j) $} is $ d_{ij}=1 $ (as shown in  Fig. \ref{originalmap1}), and the NFV delay of each function is $ 1 $ at cloud node C or cloud node E, the E2E delays of service I and service II in the above solution are $ 4 $ and $ 3 $, respectively, which satisfy the E2E latency requirements of both services.

Now, consider the case that where there is no {E2E} latency constraints, i.e., removing constraints \eqref{consdelay2funs}-\eqref{delayconstraint} from problem \eqref{mip}.
Notice that this reduces to the formulation considered in \cite{Zhang2017}.
Since the objective is to minimize the number of activated cloud nodes, the obtained solution is that both functions are processed by cloud node E:\vspace{0.05cm}\\
$\begin{array}{l}
		\text{Service I}: \text{A} \rightarrow \text{B} \rightarrow \text{E~{(providing function $ f^1 $)}} \rightarrow  \text{D},                        \\
		\text{Service II}: \text{A} \rightarrow \text{C} \rightarrow \text{E~{(providing function $ f^2 $)}} \rightarrow  \text{D}  \rightarrow \text{B}. \\
	\end{array}$\vspace{0.05cm}\\
For service II, it traverses $ 4 $ links from node A to node B with a communication delay being $ 4 $, which, pluses the NFV delay $ 1 $, obviously violates {its E2E latency constraint}.

Next, suppose that there is only one service from node A to node D with the E2E delay threshold being $ \Theta_{1}=4 $.
The considered service contains function $ f^1 $ and both of the service function rates $ \lambda_0(1) $ and $ \lambda_1(1) $ are $ 4 $.
If only a single path is allowed to transmit the traffic flow as in \cite{Woldeyohannes2018}, no solution exists for this example due to the {\emph{limited}} link capacity.
However, in sharp contrast, using our formulation \eqref{mip} with $P=2$, the traffic flow can be flexibly transmitted on multiple paths, which gives us a feasible solution as follows: first use paths $ \text{A} \rightarrow \text{B} \rightarrow \text{E} $ and $ \text{A} \rightarrow \text{C} \rightarrow \text{E} $ to route the flow from A to E where the data rates on both paths are $ 2 $; after function $ f^1 $ being processed by cloud node E, route the flow to the destination node D using the link $  \text{E} \rightarrow \text{D}$.
This toy example clearly shows the benefit of flexible routing in our proposed formulation \eqref{mip}, i.e., it has a lower requirement on the link capacities of the network to support the services.


\subsection{Simulation Results}

In this subsection, we present more simulation results to illustrate the effectiveness of the proposed formulation compared to those in \cite{Zhang2017} and \cite{Woldeyohannes2018}.

We randomly generate a network consisting of $ 6 $ nodes on a $ 100\times100 $ region in the Euclidean plane including $ 3 $ cloud nodes.
%
%
We generate link $ (i,j) $ for each pair of nodes $ i $ and $ j $ with the probability of $0.6$.
The communication delay on link $ (i,j) $ is calculated by the distance of link $ (i,j) $ over $ \bar{d} $, where $ \bar{d} $ is the average length of all shortest paths between every pair of nodes.
The cloud node and link capacities are randomly chosen in $ [6,12] $ and $ [0.5,3.5] $, respectively.
There are in total $ 5 $ different service functions: $ \{f^1, \ldots, f^5\} $.
Among the $3$ cloud nodes, $2$ cloud nodes are randomly chosen to process $ 2 $ service functions of $ \{f^1, \ldots, f^5\} $ and the remaining one is chosen to process all the service functions.
The processing delay of each function in each cloud node is randomly chosen in $ [0.8,1.2] $.
For each service $ k $, nodes $ S(k) $ and $ D(k) $ are randomly chosen from the available network nodes excluding the cloud nodes; SFC $ \mathcal{F}(k) $ is an ordered sequence of functions randomly chosen from $ \{f^1, \ldots, f^5\} $ with $ |\mathcal{F}(k)=3| $; the service function rates $ \lambda_s(k) $ are all set to $ 1 $; and the E2E delay threshold $ \Theta_k $ is set to $ 3+(6*\text{dist}_k/\bar{d}+\alpha) $ where $ \text{dist}_k $ is the length of the shortest path between nodes $ S(k) $ and $ D(k) $ and $ \alpha $ is randomly chosen in $[0,2]$.
The above parameters are carefully chosen such that the constraints of problem \eqref{mip} are neither too tight nor too loose. 

In our simulations, we randomly generate 100 problem instances for each fixed number of services and the results presented below are based on statistics from
all these 100 instances.
In problem \eqref{mip}, we choose $ P=2 $.
We use Gurobi 9.0.1 \cite{Gurobi} to solve all MBLP problems.
\begin{figure}[h]
	\centering
	\includegraphics[height=1.9in]{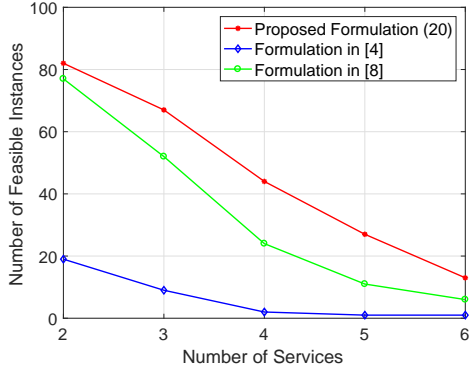}
	\caption{{Number of feasible problem instances.}}
	\label{feascases}
\end{figure}

%
Since the formulation in \cite{Zhang2017} does not explicitly take the latency constraints into consideration, the blue curve in Fig. \eqref{feascases} is obtained as follows.
We solve the formulation in \cite{Zhang2017} by changing its objective into minimizing the number of activated cloud nodes and then substitute the obtained solution into the latency constraints in \eqref{delayconstraint}: if the solution satisfies all latency constraints, we count the corresponding problem instance feasible; otherwise it is infeasible.
We can see from Fig. \ref{feascases} that the number of feasible problem instances of solving our proposed formulation \eqref{mip} is significantly larger than that of solving the formulation in \cite{Zhang2017}.
This clearly shows the advantage of our proposed formulation (i.e., it has a guaranteed E2E Latency) over that in \cite{Zhang2017}.
In addition, the flexibility of traffic routing in our proposed formulation \eqref{mip} {enables} it to also solve a larger number of problem instances than that can be solved by using the formulation in \cite{Woldeyohannes2018}.
%
%
These results further illustrate the effectiveness of our proposed formulation \eqref{mip} (as compared to those in \cite{Zhang2017} and \cite{Woldeyohannes2018}).
\begin{figure}[h]
	\centering
	\subfigure{
		\includegraphics[height=1.32in]{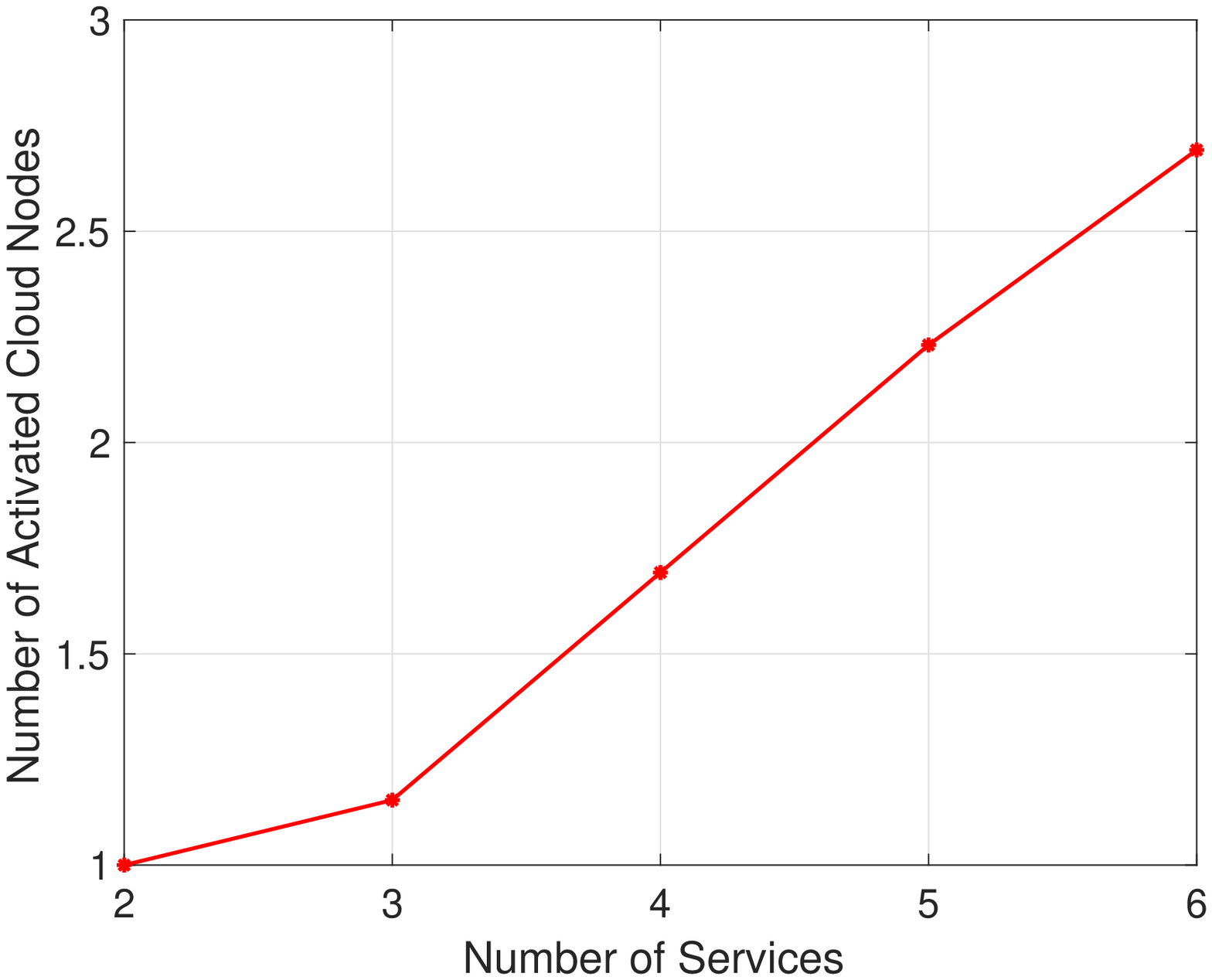}
		\label{activatednodesres1}
	}
	\subfigure{
		\includegraphics[height=1.32in]{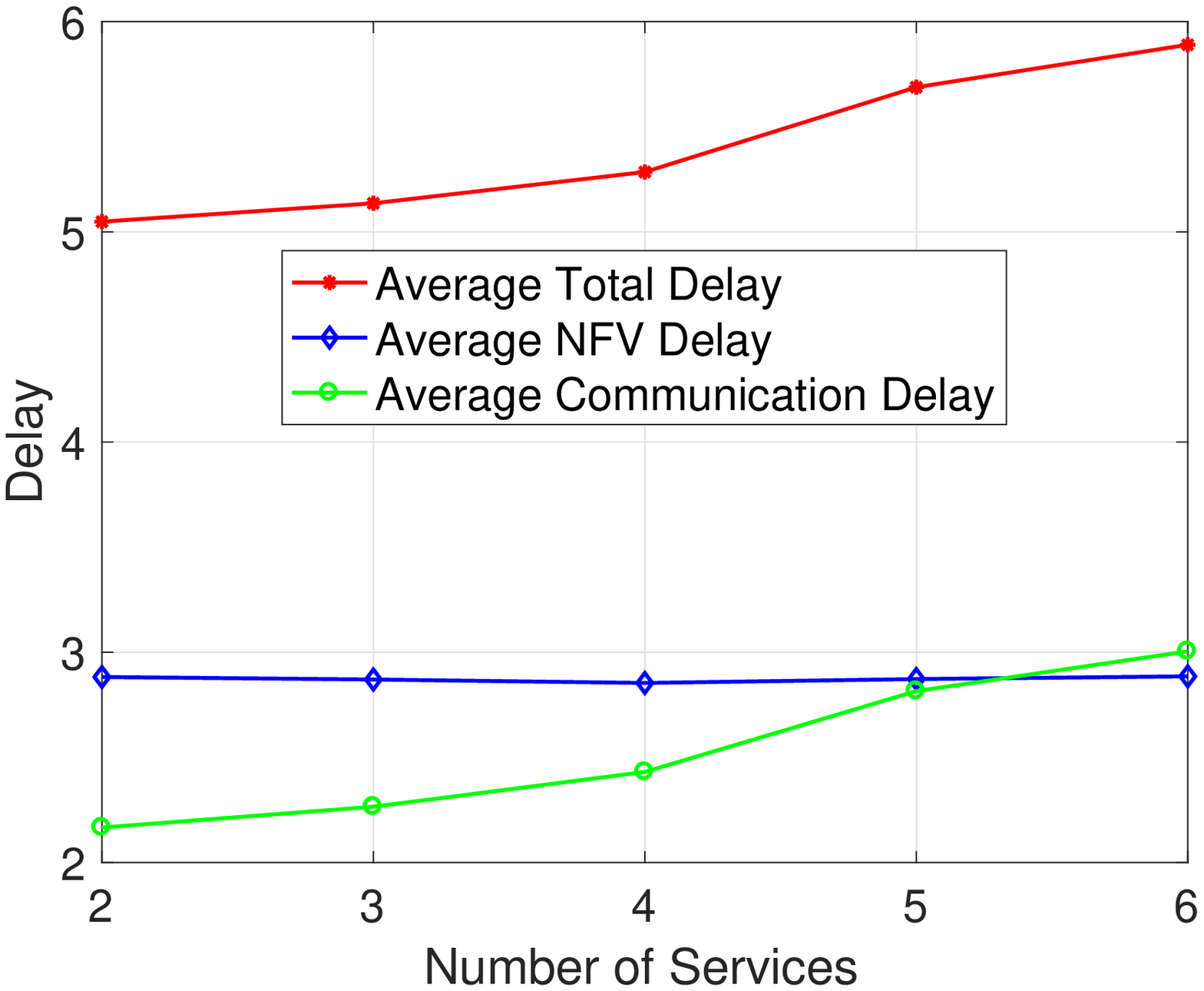}
		\label{delaysres1}
	}
	\caption{{Left: the number of activated cloud nodes; Right: the average total, NFV, and communication delays.}}
	\label{diffnflows}
\end{figure}

We now show the performance of problem \eqref{mip} versus the number of services.
%
%
Fig. 3 is obtained by averaging the results over all feasible problem instances.
We can observe from Fig. 3 that: as the number of services increases, more cloud nodes need to be activated; the average NFV delay almost keeps unchanged.
The later is due to the fact that the number of functions in all services is fixed (to be 3) and the difference of NFV delays on different cloud nodes is small.
However, the average communication delay (and the average total delay) increases rapidly.
This is mainly due to the limited link capacity.
More specifically, as the network traffic gets heavier, a traffic flow may use a path with a larger communication delay as some link's capacity (in a path with a smaller communication delay) is not enough for the data rate.

\section{Conclusions}
In this paper, we have investigated the network slicing problem that plays a crucial role in 5G and B5G networks.
We have proposed a new MBLP formulation for the network slicing problem, which can be optimally solved by the standard {solvers} like Gurobi.
Our proposed formulation minimizes the total power consumption of the whole network (equivalent to the total number of activated cloud nodes) subject to the SFC constraints and the E2E latency constraints of all services and all cloud nodes' and links' capacity constraints.
Numerical results demonstrate the advantage of our proposed formulation over the existing ones in \cite{Zhang2017} and \cite{Woldeyohannes2018}.
%



\ifCLASSOPTIONcaptionsoff
	\newpage
\fi

\bibliographystyle{IEEEtran}


%







\end{document}